\documentclass[epj]{mysvjour}
\usepackage[dvips]{graphicx}
\begin{document}
\title{Dispersion of incoherent spectral features in systems with strong electron-phonon coupling}
\titlerunning{Dispersion of incoherent spectral features
in systems with strong \dots}
\author{O. R\"osch\thanks{e-mail: \texttt{O.Roesch@fkf.mpg.de}} \and O. Gunnarsson}
\institute{Max-Planck-Institut f\"ur Festk\"orperforschung,
Postfach 800665, D-70506 Stuttgart, Germany}
\date{Received: date / Revised version: date}
\abstract{
We study (inverse) photoemission from systems
with strong coupling of doped carriers to phonons.
Using an adiabatic approximation, we develop a method for 
calculating spectra. This method is particularly simple for
systems where the electron-phonon coupling can be neglected in
the initial state, e.g., the undoped $t$-$J$ model. The theory 
then naturally explains why the electron-phonon coupling just
leads to a broadening of spectra calculated without electron-phonon 
coupling, without changing the dispersion. This is in agreement
with recent angle-resolved photoemission spectroscopy (ARPES) 
on undoped cuprates, and it supports the interpretation in terms 
of strong electron-phonon interaction. The theory also shows that
for systems with strong electron-phonon coupling in the initial
state, the result cannot in general be related to the spectrum
obtained without electron-phonon coupling.
 % end of PACS codes
} %end of abstract
\maketitle
\section{Introduction}

Angle-resolved photoemission spectroscopy
(ARPES) experiments have found evidence for strong elec\-tron-phonon 
interaction (EPI) and polaron physics in many materials like
quasi-one-dimensional conductors \cite{Perfetti,Perfetti2}, the 
manganites \cite{Dessau}, or the undoped high-$T_c$ cu\-prates 
\cite{ARPES1,ARPES2,Kyle,Kyle2}.
The spectra show an incoherent broad feature whereas the quasi-particle peak at lower binding energy is strongly suppressed.
In the case of the undoped high-$T_c$ cu\-prates,
the dispersion of that broad peak matches the
quasi-particle dispersion calculated in purely electronic models \cite{Kyle}.
For an undoped $t$-$J$ model with coupling of doped holes to optical phonons, numerical calculations of the ARPES spectra indeed showed broad features tracing the dispersion
of the quasi-particles in the original $t$-$J$ model \cite{MN}.
Similar observations in the manganites \cite{Dessau} and quasi-one-dimensional conductors \cite{Perfetti,Perfetti2} have been interpreted in
analogy with a single electron coupled to harmonic oscillators \cite{Mahan}
and a related sum-rule for the first spectral moment. With respect to the manganites also the picture of the photohole seeing a \emph{frozen} lattice has been used \cite{Perebeinos}.

In this paper we address (inverse) photoemission spectra from systems 
with strong coupling of doped carriers to phonons. We develop a theory
based on the adiabatic approximation. This theory takes a particularly
simple form if the EPI can be neglected in the initial state. This is 
the case for the undoped $t$-$J$ model and the empty or full Holstein
model. The spectrum can then be related to the spectrum of a model
without EPI, and the effect of the EPI is essentially to broaden the
spectrum. For systems where the EPI is important also in the initial
state, the theory takes a more complicated form. It can then be related 
to an average of spectra for  {\it distorted} lattices without EPI.

After giving general arguments in section \ref{general}
we discuss numerical results for the Holstein model (section \ref{sechol})
and the $t$-$J$ model with phonons (section \ref{sectj}) as illustrating examples.

\section{General considerations}\label{general}

We consider a system that is modeled by the following Hamiltonian:
\begin{equation}
H=H_{el}+H_{ph}+H_{ep}.
\end{equation}
$H_{el}$ ($H_{ph}$) describes the purely electronic (phononic) part of the model whereas
the interaction between electrons and phonons is given by $H_{ep}$.

The phonons are assumed to be harmonic in the absence of EPI and the 
system is taken to be translationally invariant, so that we can write:
\begin{equation}\label{Hphon}
H_{ph}=\sum_{{\bf q},\nu}\frac{1}{2}\left(
\Pi_{{\bf q},\nu}\Pi_{-{\bf q},\nu}+\omega_{{\bf q},\nu}^2 Q_{{\bf q},\nu}Q_{-{\bf q},\nu}\right).
\end{equation}
Here, an individual phonon mode with frequency $\omega_{{\bf q},\nu}$ has wavevector ${\bf q}$ and belongs to branch $\nu$.
Its generalized coordinate and momentum are denoted by $Q_{{\bf q},\nu}$ and $\Pi_{{\bf q},\nu}$.

The EPI couples electronic degrees of freedom to the phonon coordinates $Q_{{\bf q},\nu}$.
We assume that
this interaction vanishes for a certain electronic filling of the system which we will refer to as undoped in the following.
The completely empty or completely filled Holstein or Holstein-$t$-$J$ model are examples for such undoped systems, see
sections \ref{sechol} and \ref{sectj}.

In addition, we assume
that the phonon frequencies are small compared to the
electronic energy scales defined by $H_{el}$. This justifies
an adiabatic approximation \cite{Born} and we can first consider only $H_{el}+H_{ep}$ treating the phonon coordinates
$Q_{{\bf q},\nu}$ in $H_{ep}$ as c-numbers, i.e. as instantaneous parameters for the electronic problem. We denote the corresponding
eigenstates and eigenvalues by $|E^{N_e}_m(\vec Q)\rangle$ and $E^{N_e}_m(\vec Q)$ which are labeled by
the number of electrons $N_e$ and other quantum numbers $m$.
We use $\vec Q$ as a shorthand notation for
the set of phonon coordinates.
The phonon eigenfunctions are then obtained by solving
\begin{equation}\label{SE}
%(H_{ph}+E^{N_e}_m(\vec Q))
\left(\sum_{{\bf q},\nu}\frac{1}{2}
\Pi_{{\bf q},\nu}\Pi_{-{\bf q},\nu}
+V^{N_e}_m(\vec Q)\right)
\phi^{N_e}_{mn}(\vec Q)
=\varepsilon^{N_e}_{mn}\phi^{N_e}_{mn}(\vec Q),
\end{equation}
a Schr\"odinger equation with the effective potential
\begin{equation}
V^{N_e}_m(\vec Q)=E^{N_e}_m(\vec Q)+
\sum_{{\bf q},\nu}\frac{1}{2}
\omega_{{\bf q},\nu}^2 Q_{{\bf q},\nu}Q_{-{\bf q},\nu}.
\end{equation}
The eigenenergies are $\varepsilon^{N_e}_{mn}$ where $n$ stands for suitable phonon quantum numbers.
In our approximation
the eigenstates of $H$ are
$\langle\vec Q|\varepsilon_{mn}^{N_e}\rangle$
$=$
$\phi^{N_e}_{mn}(\vec Q)$
$|E^{N_e}_m(\vec Q)\rangle$, i.e. we assume that
the electronic states $|E^{N_e}_m(\vec Q)\rangle$
do not mix.

The ground-state in a system with $N_e$ electrons is then given by $
\langle\vec Q|\varepsilon_{00}^{N_e}\rangle=\phi^{N_e}_{00}(\vec Q)$
$|E^{N_e}_0(\vec Q)\rangle$ with ei\-gen\-energy
$\varepsilon^{N_e}_{00}$.
In case of an undoped system with $N^0_e$ electrons there is no EPI           
and the phonon wavefunction
just corresponds to the ground-state of $H_{ph}$:
\begin{equation}\label{phi00}
\phi^{N^0_e}_{00}(\vec Q)=
\prod_{{\bf q}\nu}(\omega_{{\bf q},\nu}/\pi)^{1/4}
\exp(-\omega_{{\bf q},\nu}Q_{{\bf q},\nu}^2/2).
\end{equation}
The electronic ground-state and its eigenenergy are then independent of $\vec Q$, and $\varepsilon^{N^0_e}_{00}=E^{N^0_e}_{0}+\sum_{{\bf q},\nu}
\omega_{{\bf q},\nu}/2$.

We now do (inverse) photoemission at zero temperature
by destroying (creating)
an electron with momentum ${\bf k}$ and spin $\sigma$ in the
ground-state of a system with $N_e$ electrons.
Within the adiabatic approximation this
can be described by considering the following Green's function
\begin{eqnarray}\label{green}
G^{N_e,\mp}_{{\bf k},\sigma}(z)&=&
\langle\varepsilon_{00}^{N_e}|
\psi^{\dagger}
\frac{1}{z-(H-\varepsilon^{N_e}_{00})}
\psi
|\varepsilon_{00}^{N_e}\rangle
\\
&=&\int\!d\vec Q
\int\!d\vec Q'
\phi^{N_e*}_{00}(\vec Q)
\langle E^{N_e}_0(\vec Q)|
\psi^{\dagger}\times
\nonumber\\
&&\ \times
\langle\vec Q|
\frac{1}{z-(H-\varepsilon^{N_e}_{00})}
|\vec Q'\rangle
\psi
|E^{N_e}_0(\vec Q')\rangle\phi^{N_e}_{00}(\vec Q')\nonumber
\end{eqnarray}
where $\psi=c^{(\dagger)}_{{\bf k},\sigma}$ and
$\int d\vec Q=\prod_{{\bf q},\nu}\int dQ_{{\bf q},\nu}$.

We proceed in analogy to Ref. \cite{SG} and neglect
the kinetic energy of the phonons in the resolvent in Eq. (\ref{green}).
Now $H$ is diagonal in the phonon coordinates $\vec Q$ and one half of the integrations in Eq.~(\ref{green}) can be eliminated.
This leads to the following
approximation for the Green's function \cite{SG}:
\begin{equation}\label{eq:average}
\tilde G^{N_e,\mp}_{{\bf k},\sigma}(z)=
\int\!d\vec Q
\ |\phi^{N_e}_{00}(\vec Q)|^2
g^{N_e,\mp}_{{\bf k},\sigma}(z,\vec Q)
\end{equation}
where
\begin{equation}
g^{N_e,\mp}_{{\bf k},\sigma}(z,\vec Q)=
\langle E^{N_e}_0(\vec Q)|
\psi^{\dagger}
\frac{1}{z-(H(\vec Q)-\tilde\varepsilon^{N_e}_{00})}
\psi
|E^{N_e}_0(\vec Q)\rangle.
\end{equation}
$\tilde\varepsilon^{N_e}_{00}$ is the initial state's energy without
any zero-point energy contributions arising from the phonon kinetic energy.
Finally, the corresponding spectral function is given by
\begin{eqnarray}\label{firstapprox}
\tilde A^{N_e,\mp}_{{\bf k},\sigma}(\omega)
&=&\frac{1}{\pi}\textrm{Im }\tilde G^{N_e,\mp}_{{\bf k},\sigma}(\omega\!-\!i0^+)\nonumber\\
&=&\int d\vec Q
\ |\phi^{N_e}_{00}(\vec Q)|^2
\rho^{N_e,\mp}_{{\bf k},\sigma}(\omega,\vec Q)
\end{eqnarray}
where
\begin{eqnarray}\label{firstapprox1}
&&\!\!\!\rho^{N_e,\mp}_{{\bf k},\sigma}(\omega,\vec Q)=
\sum_{m}|\langle E_m^{N_e\mp1}(\vec Q)|\psi|E_0^{N_e}(\vec Q)\rangle|^2\times\nonumber\\
&&\quad\quad\quad\quad\quad\quad\quad\quad
\times\delta(\omega-(V_m^{N_e\mp1}(\vec Q)-\tilde\varepsilon^{N_e}_{00}))
\end{eqnarray}
after expanding $\psi|E_0^{N_e}(\vec Q)\rangle$
in the adiabatic electronic basis states
$|E_m^{N_e\mp1}(\vec Q)\rangle$.

Equations (\ref{firstapprox}-\ref{firstapprox1})
will turn out to be the key formula for interpreting
ARPES spectra of undoped systems. To see this we observe that
$\rho^{N_e,\mp}_{{\bf k},\sigma}(\omega,\vec Q)$ is the spectral function
of the system without EPI for a given lattice distortion $\vec Q$.
If we assume that $V_0^{N_e}(\vec Q)$ has a non-degenerate absolute
minimum at $\vec Q_{min}$ the corresponding ground-state
phonon wave-function will be localized around this point in coordinate space.
If we approximate $|\phi^{N_e}_{00}(\vec Q)|^2\approx\delta(\vec Q-\vec Q_{min})$ we find that the spectrum corresponds to the spectrum
one obtains for the system with  a \emph{frozen} distortion $\vec Q_{min}$
in which there is no EPI.
Analogously, in case of more than one (quasi-)degenerate minima of $V_0^{N_e}(\vec Q)$ we have to take the (weighted) superposition of
the spectra corresponding to the respective distortions.
If we take into account the finite width of
$|\phi^{N_e}_{00}(\vec Q)|^2$ it follows from Eq. (\ref{firstapprox})
that the spectral features are broadened due to the
$\vec Q$-dependence of
$V_m^{N_e\mp1}(\vec Q)$.
We will consider a specific example in section \ref{sechol}.

This analysis leads to our main conclusion.
For the undoped system it follows from Eq. (\ref{phi00}) that $\vec Q_{min}=0$.
Consequently, the spectrum is just the broadened
spectrum of the same system without EPI ($H_{ep}=0$).
The dispersion of the ($H_{ep}=0$)-quasi-particle peak
shows up in the ${\bf k}$-dependence of
the broadened peak in the low binding energy part of the spectra.
This approach is particularly useful for the undoped system
as it allows statements about the spectrum for $H_{ep}\neq0$
from the knowledge of the spectrum for $H_{ep}=0$. For the doped system
$\vec Q_{min}$ is typically non-zero if the EPI is strong. The spectrum
$\rho^{N_e,\mp}_{{\bf k},\sigma}(\omega,\vec Q_{min})$ can then be
very different from the spectrum for $\vec Q=0$. Therefore,
even if the ($H_{ep}=0$)-spectrum is known,
in general no information about the spectrum of the system with
EPI can be deduced. In this context we notice that
there may be several degenerate minima at $\vec Q_{min}\neq0$
such that the ground-state has an undistorted lattice in the
sense of a vanishing expectation value of $\vec Q$. The spectrum for $H_{ep}\neq0$, nevertheless, corresponds to a superposition
of the spectra for $\vec Q$-values around the minima $\vec Q_{min}\neq0$.

The numerical calculation of spectral functions using the approximation in Eq. (\ref{firstapprox}) can be quite efficient compared to other methods typically used to obtain spectra.
In general, it is much easier to obtain the spectrum $\rho^{N_e,\mp}_{{\bf k},\sigma}(\omega,\vec Q)$ for a system without EPI but a given distortion ${\bf Q}$
than for a system with EPI. If the initial state's phonon wavefunction $\phi^{N_e}_{00}(\vec Q)$ is known (e.g. in the case of an undoped system) its square can be used as a weight function in a Monte Carlo integration over the phonon coordinates in Eq. (\ref{firstapprox}). The numerical effort is independent of the strength of the EPI. On the other hand, if e.g. exact diagonalization is used to obtain directly the ARPES spectrum of a system
with EPI, calculations become computationally very demanding with increasing coupling strength as the truncated phonon Hilbert space grows larger and larger.

In order to improve the present approximation we have to include also the kinetic energy of the phonons
in the resolvent of the Hamiltonian. Then, $\psi|E_0^{N_e}(\vec Q)\rangle$ in Eq.~(\ref{green}) must be expanded with respect to both
electronic and phononic basis functions
in the adiabatic approximation. From this, one obtains
the following expression for the spectral function:
\begin{eqnarray}
\label{apm}
&&\!\!A^{N_e,\mp}_{{\bf k},\sigma}(\omega)\!=\!
\sum_{m,n}
\delta(\omega-(\varepsilon_{mn}^{N_e\mp1}-\varepsilon^{N_e}_{00}))
\times\\
&&\quad\times\left|
\int\!d\vec Q
\ \langle E^{N_e\mp 1}_m(\vec Q)|\psi|E^{N_e}_0(\vec Q)\rangle\phi_{mn}^{N_e\mp1}(\vec Q)
\phi_{00}^{N_e}(\vec Q)
\right|^2.\nonumber
\end{eqnarray}
Each eigenstate of $H$ in the $(N_e\mp1)$-electron sector
represents a possible final state. It contributes to the spectrum at its eigenenergy (shifted by the ground-state energy of the system
with $N_e$ electrons). The intensity is proportional to the squared overlap of final and initial state, i.e. the ground-state of the system
with $N_e$ electrons plus an additional hole or electron.
Two conditions must be fulfilled for this overlap to be large.\footnote{In the following discussion we assume a non-degenerate minimum of $V_0^{N_e}(\vec Q)$ for simplicity. The arguments can be easily generalized to the case of (quasi-)degenerate minima.}
\newline
i) As the initial phonon
wavefunction
$\phi_{00}^{N_e}(\vec Q)$
is localized around the minimum
$\vec Q_{min}$ of $V_0^{N_e}(\vec Q)$
and has no nodes,
the final phonon wavefunction
$\phi_{mn}^{N_e\mp1}(\vec Q)$
must have a large
and slowly varying amplitude in this region, too. This will be the
case for final states with energies
$\varepsilon^{N_e\mp1}_{mn}\approx V^{N_e\mp1}_m
(\vec Q_{min})$.
For smaller energies the region
around $\vec Q_{min}$ is classically forbidden and the amplitude of the final pho\-non wavefunction becomes exponentially
suppressed, where\-as for larger energies the kinetic energy increases and the wavefunction oscillates faster. In both cases the integrated
overlap of initial and final phonon wavefunction becomes smaller again.
\newline
ii) The electronic
matrix element
$\langle E^{N_e\mp 1}_m(\vec Q)|\psi|E^{N_e}_0(\vec Q)\rangle$
must be large around $\vec Q_{min}$. It is sufficient to consider its value only in this region as the initial phonon wavefunction is small elsewhere.

Altogether this leads to the following picture:
For a system without EPI
but with a given lattice distortion $\vec Q_{min}$
the spectrum consists
of $\delta$-functions at the energies $E^{N_e\mp 1}_m(\vec Q_{min})$.
If the EPI is switched on, spectral features with large intensities will still
appear at similar energies and with similar relative weight but they will be broadened by phonon sidebands. The quasiparticle's dispersion and weight, however, can be strongly altered by the EPI.
In general, the effective phonon potential
$V_0^{N_e\mp1}(\vec Q)$ corresponding to the electronic ground-state
in the system with $N_e\mp1$ electrons
has minima at $\vec Q\neq\vec Q_{min}$.
The ground-state phonon wavefunction
$\phi_{00}^{N_e\mp1}(\vec Q)$
is localized around these minima.
Consequently, there is only little overlap with the phonon wavefunction
in the initial state which peaks around $\vec Q_{min}$ and in the spectrum the peak lowest in binding energy has only very small weight.

It is interesting to discuss the problem above in terms of a sum-rule
concerning the first moment (center of gravity) of the spectrum.
For the undoped system with $N^0_e$ electrons that has no phonons excited in the initial state,
one can show under rather general assumptions that the first moment of the (inverse) photoemission spectrum
does not depend on the strength of the EPI.
Let us consider the case when $A^{N^0_e,\mp}_{{\bf k},\sigma}(\omega)$ has only one peak for a given ${\bf k}$ in the absence of EPI, e.g. if there
is only one band and
no electron-electron interaction.
Turning on the EPI then broadens the peak, but due to the sum-rule the interaction is normally not expected to drastically change the spectrum in other respects.
If the electrons interact with themselves, however,
already for systems without EPI
$A^{N^0_e,\mp}_{{\bf k},\sigma}(\omega)$ usually has several peaks for a given ${\bf k}$ and the first moment does not correspond to the quasi-particle energy.
Then, the sum-rule is not able to tell us how the different peaks are broadened and cannot be used to argue for prominent features in the spectrum dispersing
approximately like the quasi-particles in the system without phonons.

\section{Holstein model}\label{sechol}

In the following we demonstrate the validity of the arguments given above with several examples.
First, we consider the one-dimensional $N$-site Holstein model with periodic boundary conditions for which the electronic part of the Hamiltonian just contains nearest-neighbor hopping with amplitude $t$:
\begin{equation}
H_{el}=-2t\sum_{k,\sigma}\cos(k)\ c_{k,\sigma}^{\dagger}c_{k,\sigma}^{\phantom\dagger},
\end{equation}
where $c_{k,\sigma}^{\dagger}$ creates an electron with momentum $k$ and spin $\sigma$. The electrons are coupled to dispersionless phonons with frequency $\omega_{ph}$ as described by
\begin{equation}
H_{ep}=\frac{g}{\sqrt{N}}\sum_{q,k,\sigma}
\sqrt{2\omega_{ph}}Q_qc_{k+q,\sigma}^{\dagger}c_{k,\sigma}^{\phantom\dagger}.
\end{equation}
The strength of the interaction is given by $g$. The $(q=0)$-phonon mode couples to the total number of electrons, $N_e$. We can therefore write $H=H_1+H_2$, where
\begin{equation}
H_1=N_e\frac{g}{\sqrt{N}}
\sqrt{2\omega_{ph}}Q_0+
\frac{1}{2}\left(
\Pi_0^2+\omega_{ph}^2 Q_0^2\right)
\end{equation}
can be solved exactly \cite{Mahan}. The spectral function $A^{N_e,\mp(2)}_{k,\sigma}(\omega)$ obtained for $H_2$ only needs to be convoluted by the known result for $H_1$,
\begin{equation}
\label{A1}
A^{N_e,\mp(1)}(\omega)=e^{-\alpha}\sum_{l=0}^{\infty}
\frac{\alpha^l}{l!}
\delta(\omega+(1\mp2N_e)\alpha\omega_{ph}-l\omega_{ph}),
\end{equation}
where $\alpha=(g/\omega_{ph})^2/N$, in order to get $A^{N_e,\mp}_{k,\sigma}(\omega)$ as defined in Eq.~(\ref{apm}).
We will therefore restrict our discussion to $A^{N_e,\mp(2)}_{k,\sigma}(\omega)$ in the following.

\begin{figure}
\includegraphics[width=8cm]{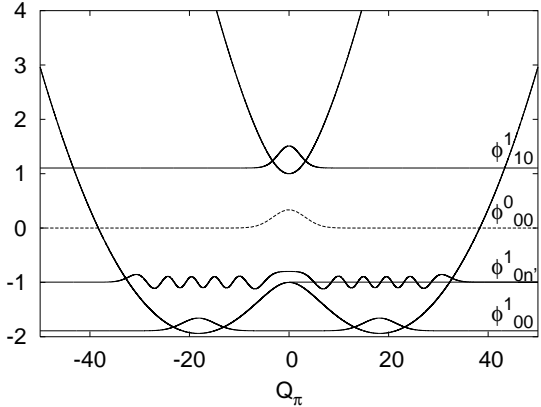}
\caption{\label{fig1}Effective potentials in the two-site Holstein model with one electron as functions of the phonon coordinate $Q_{\pi}$ for $t=1$, $\omega_{ph}=0.1$, $g=0.6$. Some selected phonon wavefunctions are also shown, see text.}
\end{figure}

\begin{figure}
\includegraphics[width=8cm]{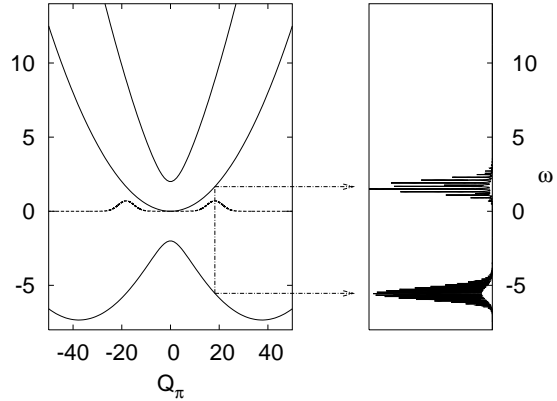}
\caption{\label{figb}Effective potentials in the two-site Holstein model with two electrons with opposite spin as functions of the phonon coordinate $Q_{\pi}$ for $t=1$, $\omega_{ph}=0.1$, $g=0.6$
are shown in the left panel together with the phonon wavefunction $\phi^{N_e=1}_{00}$ (dotted line). Right panel: $A^{N_e=1,+(2)}_{k=0,\sigma}(\omega)$ rotated by 90$^{\circ}$.}
\end{figure}

We specifically consider a two-site system ($N=2$) \cite{Ranninger} and calculate the inverse photoemission spectrum for creating an electron in both the empty (undoped) system and the system that already contains an electron of opposite spin. There is then only one phonon coordinate $Q_{\pi}$ in $H_2$ which we treat as a parameter in solving the part of $H_2$ coming from $H_{el}+H_{ep}$. In the one-electron sector one obtains the two eigenenergies
\begin{equation}
\label{effpot}
E^{N_e=1}_{0/1}=\mp\sqrt{t^2+\omega_{ph}g^2Q^2_{\pi}},
\end{equation}
whereas there are four eigenergies in case of two electrons with opposite spin:
\begin{equation}
\label{effpot1}
E^{N_e=2}_{0/3}=\mp2\sqrt{t^2+\omega_{ph}g^2Q^2_{\pi}}
\textrm{,\quad}
E^{N_e=2}_{1/2}=0.
\end{equation}
For numerical calculations we choose $t=1$, $\omega_{ph}=0.1$, and $g=0.6$. As $\omega_{ph}/t$ is small our adiabatic approximation is justified. The effective potentials one obtains by adding the harmonic potential $\omega_{ph}^2Q_{\pi}^2/2$ to the eigenenergies
in Eqs. (\ref{effpot}) and (\ref{effpot1}) are shown in Figs. \ref{fig1} and \ref{figb}, respectively.

\begin{figure}
\includegraphics[width=8cm]{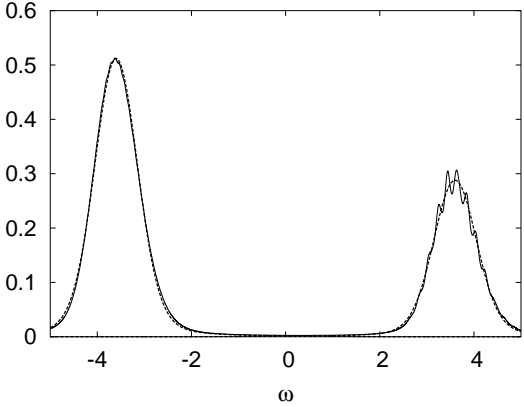}
\caption{\label{figc}
Spectrum for adding an electron to a two-site Holstein model with one electron of opposite spin ($t=1$, $\omega_{ph}=0.1$, $g=0.6$). The
approximation
$\tilde A^{N_e=1,+(2)}_{k=0,\sigma}(\omega)$
(Eq. (\ref{worst})) is shown as dashed line together with the spectrum
obtained from exact diagonalization (thin line). Both spectra have been convoluted with a Lorentzian (FWHM=0.1).}
\end{figure}

We first consider the approximation when the kinetic energy of the phonons is neglected in the resolvent of the Hamiltonian
and for which the spectral function is given by Eq. (\ref{firstapprox}).
The neglected terms are proportional to $\omega_{ph}$ so we cannot expect to resolve fine-structure in the spectra on that order. But it turns out that this approximation still describes the overall broadening correctly on a larger scale proportional to $\sqrt{\omega_{ph}}$ \cite{SG}.
The phonon wavefunction in the initial state is known exactly for the undoped system , see Eq. (\ref{phi00}). Here
\begin{equation}
\phi^{N^0_e=0}_{00}(Q_{\pi})
=
(\omega_{ph}/\pi)^{(1/4)}
\exp(-\omega_{ph}Q_{\pi}^2/2).
\end{equation}
In the system with one electron the lowest effective potential $V^{N_e=1}_0(Q_{\pi})$
has two minima at $\overline Q_{\pm}$ around which we can treat it
as harmonic potential
 with renormalized phonon frequency $\overline\omega=\sqrt{(\partial^2V_0^{N_e=1}/\partial Q_{\pi}^2)|_{\overline Q_{\pm}}}$. This leads to the approximation
\begin{equation} \phi^{N_e=1}_{00}(Q_{\pi})\approx\frac{1}{\sqrt{2}}\sum_{i=\pm}\left(\frac{\overline\omega}{\pi}\right)^{1/4}\exp(-\overline\omega(Q_{\pi}-\overline Q_i)^2/2).
\end{equation}
Following Ref. \cite{SG},
we can expand the argument of the $\delta$-functions in Eq. (\ref{firstapprox}) up to first order in $Q_{\pi}$ around $\overline Q_{\pm}$. If we further assume that the electronic matrix elements vary only weakly around $\overline Q_{\pm}$ the integration over $Q_{\pi}$ in Eq. (\ref{firstapprox}) can be eliminated and we obtain the following result:
\begin{equation}\label{worst}
\tilde A^{N_e=1,+(2)}_{k,\sigma}\!(\omega)\!\approx\!
\sum_m\!
\left.|\langle E_m^{2}|c^{\dagger}_{k,\sigma}|E_0^{1}\rangle|^2\right|_{\overline Q_{\pm}}
\!\!\sqrt{\frac{\overline\omega}{\pi b_m^2}}
e^{-\frac{\overline\omega}{b_m^2}(\omega-a_m)^2}
\end{equation}
where $a_m=V_m^{2}|_{\overline Q_{\pm}}-\tilde\varepsilon^{1}_{00}$ and $b_m=(\partial V_m^{2}/\partial Q_{\pi})|_{\overline Q_{\pm}}$. This approximation is shown in Fig. \ref{figc} for $k=0$ together with the spectrum obtained from exact diagonalization. The agreement
with the Gaussian lineshape predicted by Eq. (\ref{worst})
is very good. Only to resolve the fine-structure on a scale given by the phonon frequency $\omega_{ph}$ one has to go beyond the present approximation.
As indicated schematically by the arrows in Fig. \ref{figb} the spectrum can indeed be understood as the spectrum of electrons
in a system without EPI but a given distortion
$\overline Q_{\pm}$. The broadening is due to the finite width of the phonon wavefunction in the initial state. There is no structure in the spectrum arising from the highest effective potential as the corresponding electronic matrix element vanishes.

\begin{figure}
\includegraphics[width=8cm]{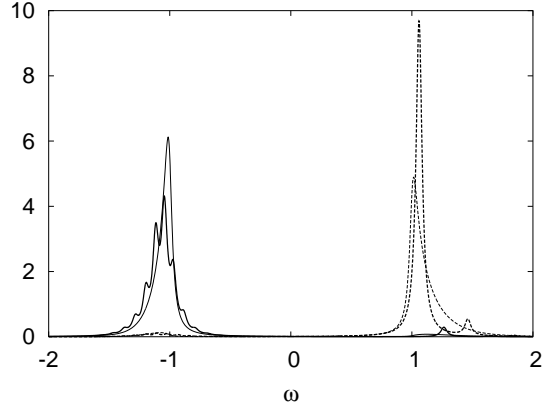}
\caption{\label{figa}
Spectra for adding an electron to an empty two-site Holstein model ($t=1$, $\omega_{ph}=0.1$, $g=0.6$).
The approximation
$\tilde A^{N^0_e=0,+(2)}_{k,\sigma}(\omega)$
(Eq. (\ref{firstapprox}), thin lines) and results from exact diagonalization (bold lines) are shown for both $k=0$ (solid lines) and $k=\pi$ (dashed lines). All spectra include a Lorentzian broadening (FHWM=0.06 eV).}
\end{figure}

In the case of inverse photoemission from the empty system the phonon wavefunction in the initial state is localized around $Q_{\pi}=0$.
The slope of both effective potentials $V^{N_e=1}_{m=0,1}(Q_{\pi})$ vanishes at this point. An approximation analogous to Eq. (\ref{worst})
would therefore result in $\tilde A^{N^0_e=0,+(2)}_{k,\sigma}(\omega)
$$=$$\rho^{N^0_e=0,+}_{k,\sigma}(\omega,Q_{\pi}=0)$, i.e. the unbroadened spectrum of the undistorted system without EPI. If we evaluate Eq. (\ref{firstapprox}) without any further approximation we obtain the spectra shown in Fig. \ref{figa}.
The comparison with results from exact diagonalization shows that
this approximation cannot reproduce the fine-structure and
for $k=0$ does not give spectral weight above $\omega=-1$ (except from broadening),
%predicts wrongly a strong asymmetry of the main peak for $k=0$,
but it still
gives the right order for the broadening of the peaks.

\begin{figure}
\includegraphics[width=8cm]{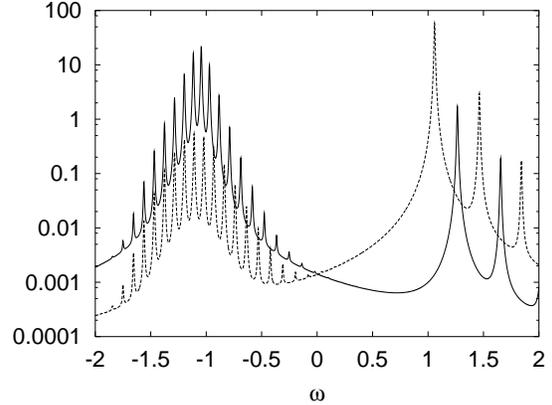}
\caption{\label{fig2}
Spectra $A^{N^0_e=0,+(2)}_{k,\sigma}(\omega)$ for adding an electron to an empty two-site Holstein model ($t=1$, $\omega_{ph}=0.1$, $g=0.6$)
as given by Eq. (\ref{apm}). The spectra for both $k=0$ (solid line) and $k=\pi$ (dashed line) are shown with Lorentzian broadening (FWHM=$0.01$). Observe the logarithmic intensity scale.}
\end{figure}

In order to understand also details of the spectra we have to use
Eq. (\ref{apm}) which results from making no approximation other than the initial adiabatic one.
One has to solve Eq. (\ref{SE}) for each effective potential to obtain the phonon eigenfunctions needed in Eq. (\ref{apm}).
The resulting inverse photoemission spectra for creating an electron with momentum $k=0$ or $k=\pi$ in the empty system are shown in Fig. \ref{fig2}. They are practically indistinguishable from those we obtained using exact diagonalization. This shows that the adiabatic approximation works very well for the chosen parameters.
For the electronic matrix element in Eq. (\ref{apm}) one needs
the $Q_{\pi}$-dependent electronic eigenstates corresponding to
the eigenenergies in Eq. (\ref{effpot}):
\begin{equation}
\label{states}
|E^{N_e=1}_{0/1}\rangle\!=\!\mathcal{N}_{0/1}\left(\!\!\left(\!t\pm\sqrt{t^2+\omega_{ph}g^2Q^2_{\pi}}\right)|0\rangle
\!-\!\sqrt{\omega_{ph}}gQ_{\pi}|\pi\rangle\!\right)
\end{equation}
with
\begin{equation}
\mathcal{N}_m=
\frac{\left(\textrm{sign}(Q_{\pi})\right)^{\alpha}}
{\sqrt{\left(t+(-1)^m\sqrt{t^2+\omega_{ph}g^2Q^2_{\pi}}\right)^2
+\omega_{ph}g^2Q_{\pi}^2}},
\end{equation}
where $|0\rangle$ and $|\pi\rangle$ denote the $k=0$ and $k=\pi$ one-electron states and $m=0,1$.

Let us consider e.g. the case where we create
an electron with momentum $k=0$ in the empty system.
The initial electronic state is then $|0\rangle$.
The initial state's phonon wavefunction
$\phi_{00}^{^{N^0_e=0}}$ (dashed line in Fig.~\ref{fig1})
is a bell-shaped Gaussian centered around $Q_{\pi}=0$.
This effectively limits the integration over $Q_{\pi}$
in Eq.~(\ref{apm}) to a small region around the origin.
Here $|E_0^{N_e=1}\rangle\approx|0\rangle$ and $|E_1^{N_e=1}\rangle\approx|\pi\rangle$.
Therefore, the relevant electronic matrix element in this example is
$\langle E_0^{N_e=1}|0\rangle$. It has even parity with respect to
$Q_{\pi}$. So, according to Eq.~(\ref{apm}) final states
with the electronic configuration $|E_0^{N_e=1}\rangle$
and a corresponding even-parity phonon wavefunction
$\phi^{N_e=1}_{0n}$ that strongly overlaps with $\phi_{00}^{N^0_e=0}$
give rise to large spectral intensity. We show in Fig.~\ref{fig1}
the even-parity phonon wavefunction $\phi^{N_e=1}_{0n'}$ with the largest overlap offset along the ordinate by its eigenergy. It has a sizable and slowly varying amplitude around $Q_{\pi}=0$ because its eigenergy is close
to the local value of the effective potential which at
$Q_{\pi}=0$ equals the ($k=0$)-eigenenergy $-t$ of the system without EPI.
Therefore, the peak with largest weight appears around this energy in the spectrum (solid line in Fig.~\ref{fig2}). The sidepeaks arise from final states with even-parity phonon wavefunctions with lower or higher eigenenergies whose overlap with $\phi_{00}^{N^0_e=0}$ decreases.
Figure~\ref{fig1} also shows the even parity ground-state phonon wavefunction $\phi_{00}^{N_e=1}$ in the double-well potential.
Clearly, its overlap with $\phi_{00}^{N^0_e=0}$ is very small
leading to a strongly suppressed quasi-particle peak in $A^{N^0_e=0,+(2)}_{k=0,\sigma}$.
Because of the large dimensionless EPI constant $\lambda=g^2/(\omega_{ph}t)=3.6$ we are well in the polaronic regime.

If, on the other hand, an electron with momentum $k=\pi$
is created in the empty state one finds using similar arguments as before
that final states leading to a large spectral intensity must have the electronic configuration $|E_1^{N_e=1}\rangle$. Their phonon wavefunction
$\phi^{N_e=1}_{1n}$ must strongly overlap with $\phi_{00}^{N^0_e=0}$ and be of even-parity. In this case the lowest energy phonon wavefunction $\phi_{10}^{N_e=1}$ in the upper effective potential (shown in Fig.~\ref{fig1}) has the largest overlap because its eigenenergy is closest to $E_1^{N_e=1}(Q_{\pi}=0)=+t$.
Therefore, $A^{N^0_e=0,+(2)}_{k=0,\sigma}(\omega)$ (dashed line in Fig.~\ref{fig2}) shows a prominent peak at $\omega\approx+t$.

As $\langle E_1^{N_e=1}|0\rangle$ ($\langle E_0^{N_e=1}|\pi\rangle$)
only vanishes completely at $Q_{\pi}=0$ the spectrum
for $k=0$ ($k=\pi$) also shows weak structures
around $\omega=+t$ ($\omega=-t$) where the coupling is now
to phonon wavefunctions of odd parity.
The density of coupling phonon states is different
around $\omega=-t$ and $\omega=+t$. E.g., the fact that the upper effective potential has a minimum around $Q_{\pi}=0$ results in an asymmetric shape of $A^{N^0_e,+(2)}_{k,\sigma}(\omega)$ around $\omega=+t$
as no phonon eigenstates in this effective potential can have eigenenergies below $+t$.

\section{$\mbox{\boldmath $t$}$-$\mbox{\boldmath $J$}$ model with phonons}\label{sectj}

As a second example we study the one-dimensional $N$-site Holstein-$t$-$J$ model with periodic boundary conditions.
This model also includes electron-electron interactions.
The electronic part of $H$ is given by the usual $t$-$J$ Hamiltonian
\begin{eqnarray}
H_{el}&=&-t\sum_{i,\sigma}\left(\tilde c_{i,\sigma}^{\dagger}\tilde c_{i+1,\sigma}^{\phantom\dagger}+h.c.\right)\\
&&\quad+J\sum_{i}\left({\bf S}_i\cdot{\bf S}_{i+1}-\frac{n_in_{i+1}}{4}\right)\nonumber,
\end{eqnarray}
where $\tilde c_{i,\sigma}^{\dagger}$ creates an electron with spin $\sigma$ on site $i$ if this site was previously empty,
$n_i=\sum_{\sigma}\tilde c_{i,\sigma}^{\dagger}\tilde c_{i,\sigma}^{\phantom\dagger}$,
and ${\bf S}_i$ is a spin-$\frac{1}{2}$ operator. Besides the hopping $t$ there is also an exchange coupling parameter $J$.
As in the Holstein-model we consider an interaction with dispersionless phonons where the coupling is now to empty sites (holes):
\begin{equation}
\label{tjint}
H_{ep}=\frac{g}{\sqrt{N}}\sum_{q,j}\sqrt{2\omega_{ph}}Q_q(1-n_j)e^{iqj}.
\end{equation}
The system with one electron per site corresponds to the undoped case where the EPI vanishes. The $(q=0)$-phonon mode can be treated separately again, the only difference being that the coupling is
now proportional to the total number of empty sites $N_h$, not to the total number of electrons $N_e$.

\begin{figure}
\includegraphics[width=8cm]{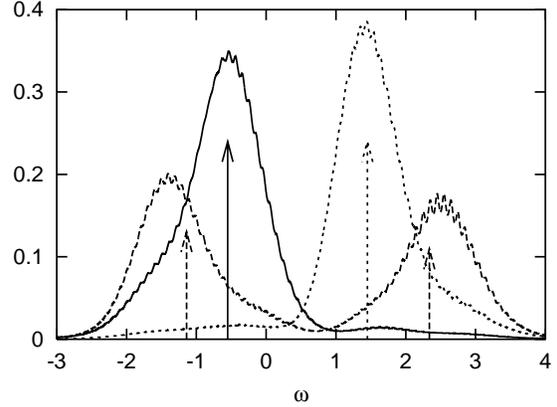}
\caption{\label{fig3}Spectral functions $A^{N^0_h=0,-}_{k,\sigma}(\omega)$ for creating a hole in the undoped 4-site Holstein-$t$-$J$ model
($t=1$, $J=0.3$, $\omega_{ph}=0.1$, $g=0.8$).
$k=0$: solid line, $k=\pm\pi/2$: dashed line, $k=\pi$: dotted line (Lorentzian broadening:
FWHM=$0.01$).
Arrows show positions and weights of corresponding peaks for $g=0$.}
\end{figure}

\begin{figure*}
(a)
\includegraphics[width=8cm]{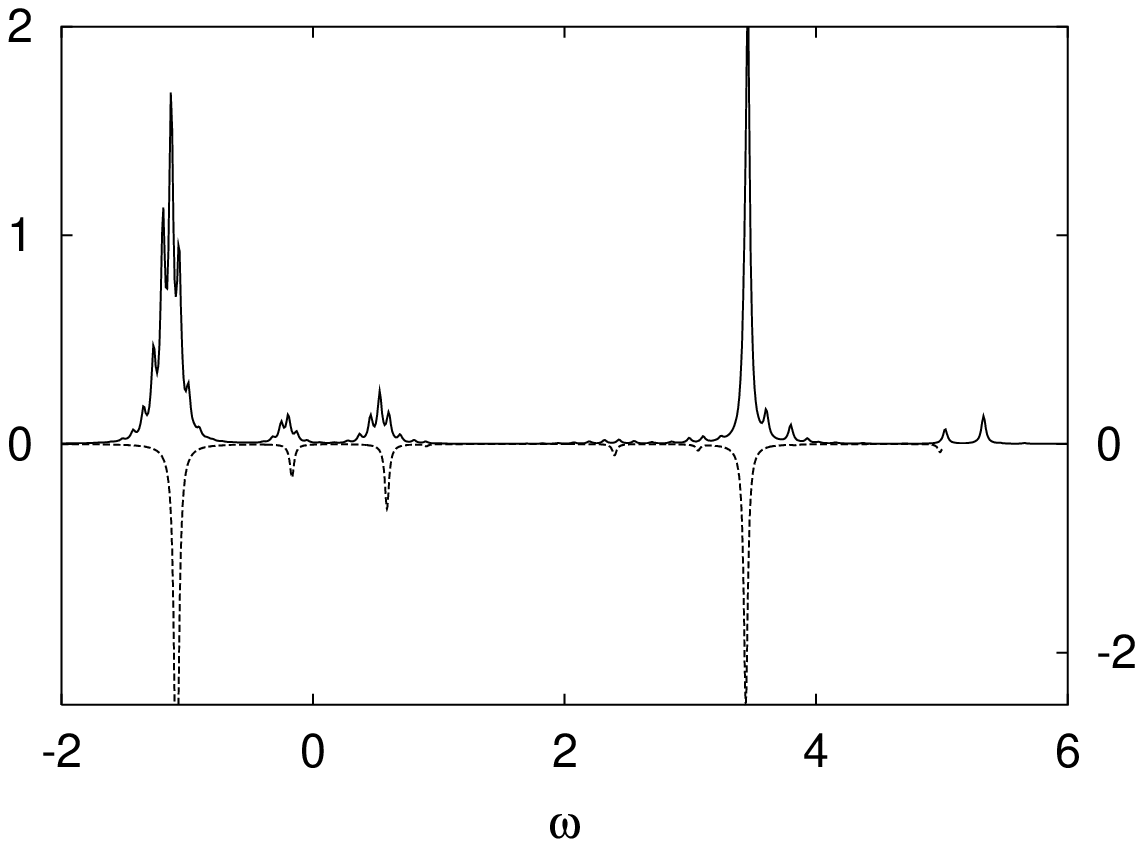}
(b)
\includegraphics[width=8cm]{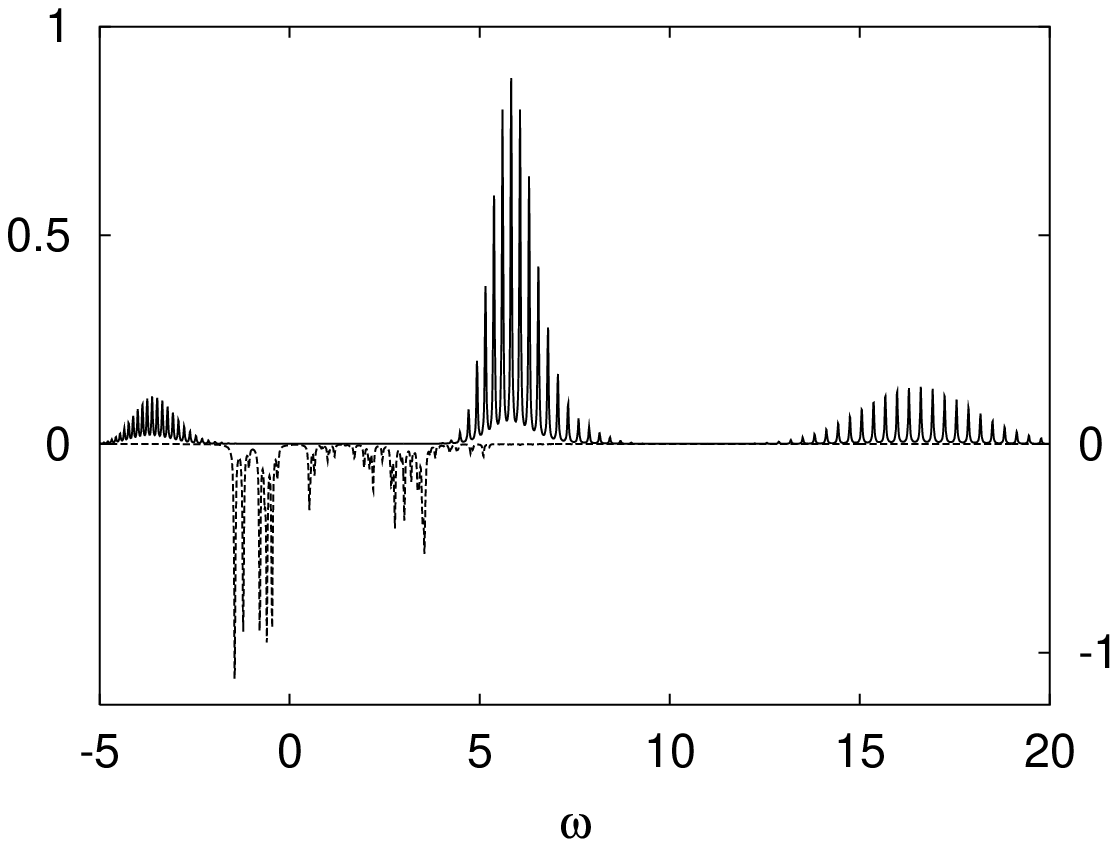}
\caption{\label{fig4}Inverse photoemission spectra
(${\bf k}=(3,1)\pi/5$)
for the 10-site $t$-$J$ model with one interacting phonon mode ($t=1$, $J=0.4$, $\omega_{ph}=0.1$, $g_{(\pi,\pi)}=1.6$):
(a) from the undoped system, (b) from the 10\%-doped system.
The corresponding spectra for systems without EPI are shown by dashed lines with their amplitude flipped for clarity.
A Lorentzian broadening of FWHM=$0.04$ has been applied.}
\end{figure*}

For numerical calculations we consider a 4-site system with $t=1$, $J=0.3$, $\omega_{ph}=0.1$, and $g=0.8$.
The photoemission spectra for destroying an electron with momentum $k$ and spin $\sigma$ were obtained using
exact diagonalization with up to 200 phonons per basis state for solving the problem without the $(q=0)$-mode
and subsequent convolution with $A^{N^0_h=0,-(1)}(\omega)$ (Eq.~(\ref{A1}) with
$N_h=0$ instead of $N_e$).
The results are displayed in Fig.~\ref{fig3} together with arrows indicating the peaks in the corresponding spectra for $g=0$.
Without EPI there is only one peak both for $k=0$ (at $-0.55$) as well as for $k=\pi$ (at $1.45$).
For $k=\pm\pi/2$ the spectrum has two peaks at $-1.139$ and $2.339$ as a result of the electron-electron interaction.

Again, the dispersion found in the system without EPI is traced quite accurately by a broad peak in the
case of strong EPI. The spectrum for $k=\pm\pi/2$ also illustrates our comments at the end of section \ref{general} on the sum-rule concerning the first spectral moment.
Since the spectrum has two peaks for $k=\pm\pi/2$, the sum rule cannot tell us how the peaks are broadened. Many other spectra would also have been consistent with the sum-rule, e.g. spectra where the peaks are shifted. The arguments based on the adiabatic approximation, however, show that both peaks should be broadened
with their individual center of gravity remaining roughly unchanged
in agreement with the exact calculations.

Finally, we consider the $t$-$J$ model in two dimensions on a tilted 10-site cluster with periodic boundary conditions. To simplify
calculations we assume that the EPI is described
by Eq. (\ref{tjint}) but with a ${\bf q}$-dependent coupling constant
$g_{\bf q}$. In the following we choose $g_{(\pi,\pi)}=1.6$ and
$g_{\bf q}=0$ for all other ${\bf q}\neq(\pi,\pi)$ so that effectively there is only one phonon mode that interacts with the electrons.
The other parameters are $t=1$, $J=0.4$, and $\omega_{ph}=0.1$.
Figure \ref{fig4} shows the inverse photoemission spectra for
${\bf k}=(3,1)\pi/5$ from both the undoped (Fig. \ref{fig4}(a)) and the 10\%-doped system (Fig. \ref{fig4}(b)). The EPI
has been switched on and off (solid line vs. dashed line with flipped amplitude).

The spectra from the undoped system confirm again our general expectations from section \ref{general}. The EPI basically broadens the structures in the original spectrum. This includes the quasi-particle peak at low binding energies. In contrast, the spectrum from the doped system changes quite differently when the EPI is switched on. Although the spectrum again develops several broad features they cannot be related anymore in a simple way to the structures in the spectrum found without EPI. We have also varied ${\bf k}$ and found that the dispersion of the broad features is different from the quasi-particle dispersion in the system without EPI. According to Eq. (\ref{firstapprox}) the
spectra rather correspond to broadened versions of spectra one would obtain in a purely electronic, but distorted system.

\section{Conclusions}

We have introduced an adiabatic approximation for calculating ARPES 
spectra from systems with strong coupling of doped carriers to phonons.
The effective phonon potential for the initial state is calculated as a function of the
phonon coordinates and its minima are found. We show that the spectrum
with electron-phonon interaction (EPI) is then related to a broadened average of spectra without EPI calculated for distorted lattices corresponding to the minima of the effective potential. We also studied the additional approximation
of neglecting the kinetic energy of the phonons in the resolvent of the Hamiltonian
corresponding to the ARPES Green's function. The spectrum is then expressed
as superposition of spectra from distorted lattices without EPI, using the square of the initial state's phonon wave function
as a weight function
(see Eq. (\ref{firstapprox})).

In either form, the theory becomes particularly simple if the EPI can
be neglected in the initial state. The phonon wave function is then
centered around the undistorted lattice, and the spectrum with EPI
can be directly related to the (broadened) spectrum without EPI.
In the case of strong EPI in the initial state, the minima of the effective potential
correspond to distorted lattices. The spectrum with EPI is then
related to the (broadened) spectra of distorted lattices without
EPI. Therefore, the knowledge of the spectrum without EPI for the undistorted lattice is in general not very informative with respect to the spectrum with EPI.

Our results support the interpretation of ARPES on undoped high-$T_c$ cuprates \cite{Kyle,Kyle2} and explain why in numerical calculations \cite{MN} the quasi-particle dispersion from purely electronic models shows up almost unchanged in the dispersion of incoherent features in the spectra obtained with EPI.

\end{document}